\documentclass[12pt,preprint]{emulateapj}

\def\be{\begin{equation}}
\def\ee{\end{equation}}
\def\bea{\begin{eqnarray}}
\def\eea{\end{eqnarray}}

\usepackage{graphicx}

\begin{document}

\title{GRB 080916C and GRB 090510: the high-energy emission and the afterglow}

\author{Wei-Hong Gao$^1$, Jirong Mao$^{2,3}$, Dong Xu$^{4}$ and Yi-Zhong Fan$^{4,5,6}$ }
\affil{$^1$ Department of Physics and Institute of Theoretical Physics, Nanjing Normal University, Nanjing, 210046, China.\\
$^2$INAF Osservatorio Astronomico di Brera, via Bianchi 46, Merate 23807, Italy.\\
$^3$ Yunnan Observatory, Chinese Academy of Sciences, Kunming, 650011, China.\\
$^4$Dark Cosmology Centre, Niels Bohr Institute, University of Copenhagen, Juliane Maries Vej 30, 2100 Copenhagen, Denmark. \\
$^5$ Niels Bohr International Academy, Niels Bohr Institute, University of Copenhagen, Blegdamsvej 17, 2100 Copenhagen, Denmark.\\
$^6$ Purple Mountain Observatory, Chinese Academy of Science,
Nanjing, 210008, China.} \email{yizhong@nbi.dk (YZF)}

\begin{abstract}
We constrain the physical composition of the outflows of GRBs
080916C and 090510 with the prompt emission data and find that the
former is likely magnetic, while the latter may be baryonic. The
X-ray and optical afterglow emission of both GRBs can be reasonably
fitted using the standard external shock model but the density
profiles of the circum-burst medium are different. We also propose a
simple method to estimate the number of seed photons supposing the
GeV afterglow photons are due to the inverse Compton radiation of
external forward shock electrons. The seed photons needed in the
modeling are too many to be realistic for both events. The
synchrotron radiation of the forward shock seems able to account for
the GeV afterglow data.
\end{abstract}

\keywords{Gamma rays: bursts---Radiation mechanisms:
non-thermal---X-rays: general}

\setlength{\parindent}{.25in}

\section{INTRODUCTION}
Gamma-ray bursts (GRBs) are the most luminous explosions in the
universe. They feature extremely relativistic outflows with bulk
Lorentz factors $\sim 10^{2-3}$ and isotropic energies of
$10^{48-55}$ erg. Though their cosmological origin as well as the
relativistic movement have been firmly established, the radiation
mechanism and the outflow composition are still uncertain
\citep{Piran99,ZM04}. It is widely believed that the high-energy
emission of GRBs can shed light on these two fundamental issues (see
Fan \& Piran 2008 for a review). For example, a distinct GeV-TeV
spectrum excess can be taken as an indication evidence of a baryonic
outflow and a radiation process in addition to synchrotron (e.g.,
inverse Compton scattering) will be needed, while the absence of
such a component in most spectra may favor the magnetic outflow
model. Recently, the Fermi collaboration has released their
observation data of GRBs 080916C and 090510 \citep{Abdo09, Abdo09b}.
In this work, we examine the origins of these prompt and afterglow
GeV emission. The work is structured as follows. In Section 2, we
discuss the origin of the prompt GeV emission and the corresponding
constraint on the physical composition. In Section 3 we employ the
standard external forward shock model to interpret the X-ray and
optical afterglow data. In Section 4, we investigate the origin of
the afterglow GeV emission. Our results are summarized in Section 5
with some discussion.

\section{Prompt GeV emission of GRBs 080916C and 090510}
{\bf GRB 080916C} was a long burst with a duration $T_{90}\simeq
66~{\rm s}$ \citep{Abdo09} and was at a redshift $z\sim 4.35 \pm
0.15$\citep{Greiner09}. A few hundred high-energy photons have been
detected by the large area telescope (LAT) onboard the Fermi
satellite and three of them are above 10 GeV. The joint analysis of
the LAT and Gamma-ray Bursts Monitor (GBM) data suggests a
featureless Band spectrum in the energy range $8~{\rm keV}-10~{\rm
GeV}$ \citep{Abdo09}. A straightforward interpretation of the
spectrum is the synchrotron radiation of internal shock electrons.
Such an interpretation, if correct, demands a very large bulk
Lorentz factor $\Gamma_{\rm i}\sim 10^{3}$ of the emitting/shocked
region \citep{Abdo09, Greiner09}. In the internal shock scenario,
the fast shells should move faster and the corresponding bulk
Lorentz factor should be $\Gamma_{\rm f} \sim 5 \Gamma_{\rm i}$
otherwise the internal shock efficiency will be too low to match the
observations \citep[e.g.,][]{Piran99}. The photosphere radius of the
fast shells is $R_{\rm ph}\sim 5\times 10^{9}~{\rm
cm}~L_{54}\Gamma_{\rm f,3.7}^{-3}$ \citep{Pacz90}, where $L$ is the
total luminosity of the outflow\footnote{In this work we adopt the
convenience $Q_{x}=Q/10^{x}$ in units of cgs.}. On the other hand,
for a baryonic shell we have $\Gamma_{\rm f}\leq R_{\rm ph}/R_0 \sim
5\times 10^{3}L_{54}\Gamma_{\rm f,3.7}^{-3}R_{0,6}^{-1}$
\citep{Piran99}, where $R_0\geq 10^{6}$ cm is the size of the
central engine. So the shell becomes transparent at the late stage
of its acceleration. As a result, the thermal radiation of these
shells will be too strong to be effectively outshone by the internal
shock non-thermal emission, in disagreement with the data (Fan 2009;
see Zhang \& Pe'er 2009 for the other approach). Hence we would not
discuss the standard/unmagnetized internal shock model for this
burst.

An interesting possibility is that the prompt emission has a very
soft MeV-GeV spectrum and the GeV photons are due to the synchrotron
radiation of the external forward shock \citep{Kumar09}. Here we
outline a few potential challenges of such a model. {(1) In the
forward shock model, the variability of the radiation is determined
by the angular timescale $T_{\rm ang}$, which is $\sim t$ as long as
the edge of the emitting region is invisible \citep{Piran99}. So the
light curve should be smooth. The variability shown in the LAT data
then disfavors the forward shock emission model.} (2) For the
initial outflow expanding into the wind medium (see Section 3 for
the medium identification), strong reverse shock may form. The bulk
Lorentz factor of the shocked medium will be almost a constant
\citep{Chevalier00}. A strong reverse shock exists till $t\sim
T_{90}/2$. In such a phase, we have the magnetic field strength $B
\propto t^{-1}$, the maximum specific flux $F_{\rm \nu,max} \propto
t^{0}$, the typical synchrotron frequency $\nu_{\rm m} \propto
t^{-1}$ and the cooling frequency $\nu_{\rm c}\propto t$. Hence the
synchrotron radiation flux in LAT band can be estimated as $F_{\rm
LAT} \propto F_{\rm \nu,max} \nu_{\rm m}^{(p-1)/2}\nu_{\rm c}^{1/2}
\propto t^{(2-p)/2}$ for $h\nu_{\rm c}<100$ MeV, inconsistent with
the observation. Where $p$ is the power-law distribution index of
the accelerated electrons at the shock front \citep[see][for
extensive discussion]{Xue09}. Since the reverse shock emission has
not been detected in most GRBs and it is not clear whether the model
suffers some disadvantages, we do not take the current temporal
inconsistence as a conclusive argument. (3) To reproduce the prompt
spectrum, the forward shock emission at $t\sim 10$ s should have
$h\nu_{\rm m} \geq 300 {\rm keV}$. At such early time, the
synchrotron self-Compton radiation is in extreme Klein-Nishina
regime and the Compton parameter $Y\sim 0$. With proper parameters,
$\rm \nu_c$ can be comparable to $\nu_{\rm m}$. So the sub-MeV
spectrum can be $F_\nu \propto \nu^{1/3}$, steep enough to be
consistent with the data. However, if $\nu_{\rm m} \sim
10^{20}(t/10)^{-3/2}~{\rm Hz}$, the XRT light curve will be
$F_{\nu_{\rm x}} \propto t^0$ for $t<10^{3}$ s and the optical light
curve will be $F_{\nu_{\rm opt}} \propto t^{0}$ for $t< 10^{5}$ s.
These behaviors are very unusual and have not been detected in other
GRB afterglows so far. The lack of observation of early afterglow of
GRB 080916C, however, hampers us to test the model.

If the prompt high-energy emission of GRB 080916C was from the soft
gamma-ray emitting region, a plausible origin of the GeV photons is
the synchrotron radiation of electrons accelerated in magnetic
energy dissipation of a Poynting-flux dominated outflow
\citep{ZP09}. A disadvantage of such a scenario is the difficulty of
reproducing the hard low energy spectrum \citep{Fan09}.

{\bf GRB 090510} was a short burst at a redshift $z\sim 0.903$
\citep{Abdo09b}. The high-energy emission is much more intense than
that of GRB 080916C and shows some variability, which disfavors the
external forward shock model. In the time interval $0.5-0.6$ s, the
sub-MeV spectrum is very hard but the high energy spectrum is very
soft \citep{Abdo09b}, possibly dominated by the photosphere emission
of the baryonic shell\footnote{The temperature of the initial shell
is $T_{\rm obs}\sim 10~{\rm
MeV}~[(1+z)/2]^{-1}L_{54}^{1/4}R_{0,6}^{-1/2}$, matching the data if
$R_{0}\sim 10^{7}$ cm. Considering the un-magnetization nature of
the outflow,  such a small $R_0$ indicates a black hole as the
central engine. The outflow  was likely launched via the
neutrino-antineutrino annihilation process.}. In the time interval
$0.5-0.8$ s, the high energy spectrum gets harder and harder but the
``thermal"-like MeV component is still evident. GeV emission is
naturally produced in the IC scattering of the ``photosphere"
photons by the shocked electrons. The photosphere radius is $\sim
6\times 10^{11}~{\rm cm}~L_{54}\Gamma_{\rm sh,3}^{-3}$, where
$\Gamma_{\rm sh}$ is the bulk Lorentz factor of the shell. The
internal shocks take place at a rather larger radius $R_\gamma \sim
\Gamma_{\rm i}^{2}c \delta t/(1+z) \sim 1.5\times 10^{15}~{\rm
cm}~\Gamma_{\rm i,3}^{2}(\delta t/0.1~{\rm s})$, where $\delta t$ is
the detected variability timescale of the prompt emission. In the
comoving frame of the emitting region the seed/photosphere photons
are moving along the radial direction and are highly anisotropic. In
such a case, the strongest IC radiation is from an angle $\sim
1/\Gamma_{\rm i}$ relative to the line of sight \citep{Fan06}. The
arrival of the GeV photons will be delayed by a time $\sim \delta t$
and the GeV radiation duration will be extended, in agreement with
the observation. Below we show how to reproduce the high energy
spectrum $F_\nu \propto \nu^{-0.54}$ in time interval $0.8-0.9$ s.
If the cooling of the electrons is dominated by the prompt soft
gamma-rays with a luminosity $L_\gamma$, the cooling Lorentz factor
can be estimated by $\gamma_{\rm c,ic} \sim 5~L_{\rm \gamma,
53.3}^{-1}R_{\rm \gamma, 15} \Gamma_{\rm i, 3}^{3}$\citep{Fan08}.
Here we do not take $L_\gamma \sim 10^{52}~{\rm erg~s^{-1}}$, the
luminosity of the simultaneous soft gamma-ray emission, since in the
photosphere-internal shock model the arrival of the upscattered
photons is delayed, as already mentioned. The corresponding IC
radiation frequency $\varepsilon_{\rm c,ic} \sim \gamma_{\rm c,
ic}^2 E_{\rm p} \sim 25~{\rm MeV}~(E_{\rm p}/1~{\rm MeV})L_{\rm
\gamma, 53.3}^{-2}R_{\rm \gamma, 15.3}^{2} \Gamma_{\rm i, 3}^{6}$,
where $E_{\rm p}$ is the typical energy of the seed photons. On the
other hand $\gamma_{\rm m,i} \approx \epsilon_{\rm e,i}(m_{\rm
p}/m_{\rm e})(\Gamma_{\rm sh}-1)/3 \approx 100 (\epsilon_{\rm
e,i}/0.5)[(\Gamma_{\rm sh}-1)/0.3]$ for $p\sim 2.5$, where
$\Gamma_{\rm sh}$ is the parameter denoting the strength of the
shocks. The corresponding IC radiation frequency is
$\varepsilon_{\rm m,ic} \sim \gamma_{\rm m, i}^2 E_{\rm p} \sim
10~{\rm GeV}~(\gamma_{\rm m, i}/100)^{2}(E_{\rm p}/1~{\rm MeV})$.
The spectrum in the energy range $\sim 10~{\rm MeV}-10~{\rm GeV}$ is
$F_\nu \propto \nu^{-1/2}$, consistent with the data. We note that
in the time interval $0.9-2$ s, the soft gamma-ray emission is very
weak while the GeV emission is still strong. These delayed GeV
photons may be produced by the IC scattering of the soft gamma-rays
by the electrons accelerated by the reverse shock or by the shocks
generated in the collision of the late time ($t>0.5$ s) outflow with
the precursor outflow.

\section{The afterglow of GRBs 080916C and 090510}
\emph{GRB 080916C. }\emph{Swift} XRT started to observe this source
at about 17 hr after the Fermi trigger. In our data analysis, the
X-ray light curve can be fitted by a single power-law $F_{\nu_{\rm
x}}\propto t^{-1.30 \pm 0.07}$ for $6.1\times 10^{4}<t<1.3\times
10^{6}$ s and the XRT spectrum is $F_\nu \propto \nu^{-0.50\pm
0.16}$. The earliest optical/infrared observation started at $t\sim
26.7$ hr after the burst. The optical/NIR light curve can be well
described by $F_{\nu_{\rm opt}} \propto t^{-1.40\pm0.05}$. The
optical to X-ray spectrum is consistent with a single power law
$F_{\nu} \propto \nu^{-0.63}$ \citep{Greiner09}. These facts suggest
that the optical to X-ray afterglow emission is within the same
regime. In the standard external shock model (e.g., Zhang \&
M\'esz\'aros 2004), the slow cooling spectrum takes the form $F_\nu
\propto \nu^{-(p-1)/2}$ and the decline should be either
$t^{3(1-p)/4}$ (ISM) or $t^{(1-3p)/4}$ (wind medium). One can see
that the X-ray and optical afterglow data are in agreement with the
wind medium model for $p\sim 2.2$ (see also Zou et al. 2009).

Assuming a GRB efficiency $\epsilon \sim 0.2$, the
isotropic-equivalent kinetic energy of the outflow is
$E_{\rm k}\sim 4\times 10^{55}~{\rm ergs}$. In the wind case, the
equations that govern the forward shock emission are \citep[e.g.,][]{Yost03}
\begin{equation}
\nu_{\rm m}\approx 1.3\times 10^{14}~{\rm Hz}~{
\epsilon_{e,-1}^{2}\epsilon_{B}^{1/2}C_{p}^{2}E_{k,55.6}^{1/2}(1+z)^{1/2}t_{4.8}^{-3/2}},
\end{equation}
\begin{equation}
\nu_{\rm c}\approx1.7\times 10^{13}~{\rm Hz}~
\epsilon_{B}^{-3/2}E_{k,55.6}^{1/2}A_{\ast}^{-2}(1+z)^{-3/2}t_{4.8}^{1/2}(1+Y)^{-2},
\end{equation}
\begin{equation}
\rm F_{\nu, max}\approx 100~{\rm mJy}~
\epsilon_{B}^{1/2}E_{k,55.6}^{1/2}A_{\ast}
t_{4.8}^{-1/2}(1+z)^{3/2}D_{L,29.1}^{-2},
\end{equation}
where $C_{p}\equiv 13(p-2)/[3(p-1)]$ for $p>2.05$,
$A_{\ast}=(\dot{M}/10^{-5}M_{\odot}~{\rm yr^{-1}})[v_{\rm
w}/(10^{8}~{\rm cm~{\rm s^{-1}}})]^{-1}$ is the wind parameter,
$v_{\rm w}$ is the speed of the wind, $\dot{M}$ is the mass loss
rate \citep{Chevalier00}, and $\rm Y=[-1+\sqrt{1+4\eta \eta_{_{\rm
KN}}\epsilon_{e}/\epsilon_{B}}]/2$, $\eta\simeq \rm
min\{1,(\nu_{m}/\nu_{c})^{(p-2)/2}\}$ and $\eta_{_{\rm KN}}$ is the
factor reflecting the importance of the Klein-Nishina correction
(see the Appendix A of \citet{Fan06a} for the expression).

Since $\nu_{\rm m}$ decreases with time while $\nu_{\rm c}$
increases with time, the current afterglow data suggest that
$\nu_{\rm m}(t=10^{5}~{\rm s})\leq \nu_{\rm opt/IR}$ and
$\nu_{\rm c}(t=6\times 10^{4}~{\rm s})\geq \nu_{\rm x}\sim 10^{18}$ Hz, i.e.,
\begin{equation}
\epsilon_{e,-1}^{2}\epsilon_{B}^{1/2}\leq 5,~~~~~~~\epsilon_{B}^{-3/2}A_{\ast}^{-2}(1+Y)^{-2} \geq 7\times 10^{5}.
\end{equation}

At $t\sim 10^{5}$ s, the $K_{s}$ band flux is $\sim
3\times10^{-5}$ Jy \citep{Greiner09}, which gives us another constraint
\begin{eqnarray}
\epsilon_{e,-1}^{1.2}
\epsilon_{B}^{0.8}A_{\ast} \sim 7\times10^{-5}.
\end{eqnarray}

Substituting $Y\sim \sqrt{\epsilon_{\rm e}/50\epsilon_{\rm B}}$ (due
to the slow cooling and the Klein-Nishina correction) in Equations
(4) and (5), we have $A_*\geq 10^{-5}\epsilon_{\rm e,-1}^{2}$,
$\epsilon_{\rm B}\geq 10^{-4}\epsilon_{\rm e,-1}^{-1.3}$. Though the
shock parameters cannot be uniquely determined, we see that the
``reasonable" parameters $(\epsilon_{\rm e}, ~\epsilon_{\rm B},~A_*)
\sim (0.1,~2.5\times 10^{-3},~0.01)$ can reproduce the data.


\emph{GRB 090510.} In our data analysis, before and after the break
at $t_{\rm b}\sim 676(1+z)$ s the X-ray declines are $t^{-0.72\pm
0.08}$ and $t^{-1.89\pm0.06}$, respectively. The X-ray spectrum can
be reasonably fitted by $F_\nu \propto \nu^{-0.63\pm 0.06}$. We
reduced the UVOT data in a standard way with the aid of reduction
threads at http://www.swift.ac.uk/UVOT.shtml. The combined V-band
and white light curves show a rise since the beginning of UVOT
observation to a peak around 1000 s after the BAT trigger, which is
followed by an apparent decay leading to the optical flux lower than
the threshold of UVOT quickly. Our results are generally in
agreement with that of \citet{Pasquale09}. Within the standard
external shock model, the above data are roughly consistent with a
slow cooling ejecta expanding into the ISM for $p\sim 2$ while the
break can be interpreted as the jet effect \citep{Piran99,ZM04}. The
slowly rising optical emission may suggest that the observer's
frequency is below $\nu_{\rm m}$. In the ISM case, the equations
that govern the forward shock emission are (e.g., Sari et al. 1998;
Yost et al. 2003)
\begin{eqnarray}
\nu_{\rm c} \approx 5.2\times 10^{18} ~{\rm Hz}~
E_{\rm k,54}^{-1/2}\epsilon_{B,-4}^{-3/2}n_{0}^{-1}(1+z)^{-1/2}t_{3.1}^{-1/2}(1+Y)^{-2},
\end{eqnarray}
\begin{eqnarray}
\rm \nu_{m}=7.0\times 10^{13} Hz
E_{k,54}^{1/2}\epsilon_{B,-4}^{1/2}\epsilon_{e,-1}^{2}C_{p}^{2}(1+z)^{1/2}t_{3.2}^{-3/2},
\end{eqnarray}
\begin{eqnarray}
\rm
F_{\nu,max}=2.7\times10^{-3}~{\rm Jy}~(1+z)D_{L,28.26}^{-2}\epsilon_{B,-4}^{1/2}E_{k,54}n_{0}^{1/2},
\end{eqnarray}
please note that we have $C_{p}\simeq0.23$ for $p\sim 2$.

The conditions that $\nu_{\rm c}(t=1284~{\rm s})>\nu_{\rm x}$,
$\nu_{\rm m}(t\sim 1000~{\rm s}) \sim 5\times 10^{14}$ Hz and
$F_{\rm \nu,max} \geq 1\times 10^{-4}$ Jy \citep{Pasquale09}
yield
\begin{equation}
E_{\rm k,54}^{-1/2}\epsilon_{B,-4}^{-3/2}n_{0}^{-1}(1+Y)^{-2} \geq 0.2,
\end{equation}
\begin{equation}
E_{k,54}^{1/2}\epsilon_{B,-4}^{1/2}\epsilon_{e,-1}^{2}\sim 50,~~~\epsilon_{B,-4}^{1/2}E_{k,54}n_{0}^{1/2}\geq 0.02.
\end{equation}

The parameters $(E_{\rm k,54},~\epsilon_{\rm B,-4},~\epsilon_{\rm
e,-1},~n_{0})\sim (1,~1,~7,~0.01)$ satisfy the above constraints
(note that $Y\ll \sqrt{\epsilon_{\rm e}/\epsilon_{\rm B}}$ thanks to
the Klein-Nishina correction). The jet break time $t_{\rm b}=1284$
sec suggests a half-opening angle $\theta_{\rm
j}=6\times10^{-3}t_{3.1}^{3/8}E_{\rm
k,54}^{-1/8}\epsilon_{-0.7}^{1/8}n_{0,-2}^{1/8}$. So the true
gamma-ray energy released is $E_{\rm \gamma,jet} \simeq \theta_{\rm
j}^{2}E_\gamma/2=2\times10^{48} \rm ergs$, where $E_\gamma \sim
1.4\times 10^{53}$ erg is the isotropic-equivalent gamma-ray energy.

\section{The high energy afterglow emission}
\subsection{IC scattering in the forward shock region?}
If the high energy afterglow is due to the IC radiation of the
forward shock electrons, there is a simple method to estimate the
number of seed photons, regardless of their origin
(either the late prompt emission from the central engine or the
synchrotron radiation of the forward shock electrons). Following
\citet{Fan06}, the possibility of one seed photon being scattered
(i.e., the optical depth) in the forward shock region can be
estimated as
\begin{equation}
\tau_{\rm ISM} \sim 4.2\times
10^{-8}~E_{k,53}^{1/4}n_0^{3/4}t_3^{1/4}[(1+z)/2]^{-1/4},
\end{equation}
\begin{equation}
\tau_{\rm wind} \sim 7.3 \times
10^{-6}~A_*^{3/2}E_{k,53}^{-1/2}t_3^{-1/2}[(1+z)/2]^{1/2},
\end{equation}
respectively. With the parameters derived for GRBs 080916C and
090510, we have
\[
\tau_{\rm wind}({\rm 080916C}) \sim
10^{-9}~A_{*,-2}^{3/2}E_{k,55.6}^{-1/2}t_{2.6}^{-1/2},
\]
\[
\tau_{\rm ISM}(090510) \sim 7\times
10^{-10}~E_{k,54}^{1/4}n_{-2}^{3/4}t_1^{1/4},
\]
respectively.

If the detected high energy afterglow photons are indeed the IC
radiation of the forward shock electrons, the number
flux of the seed photons will be
\begin{equation}
{F}_{\rm seed} \sim {{F}_{\rm >100 MeV}/ \tau }.
\end{equation}

For {\bf GRB 080916C}, in the time interval $\sim 100-1400$ s (i.e.,
$\Delta t=1300$ s), ${ F_{\rm >100 MeV}\sim 7\times 10^{-6}~{\rm
ph~cm^{-2}~s^{-1}}}$ (Abdo et al. 2009a), so the number of total
seed photons is
\begin{equation}
N_{\rm se} \sim {4\pi D_{\rm L}^2 \over (1+z)^{2}} \Delta t {F}_{\rm seed}\sim 6\times 10^{64}.
\end{equation}

If most seed photons are in the X-ray band, the total energy will be
$\sim 10^{56}$ erg, which is too large to be realistic. If the seed
photons are mainly in optical/infrared band, the total energy will
be $\sim 10^{53}$ erg. Though bright infrared/optical flare can be
produced in the afterglow phase by the prolonged activity of the
central engine (for example, the infrared flare detected in GRB
080129; \citet{Greiner09a, Gao09}), it is clear that such events are
very rare. So we think this kind of model is less likely.

For {\bf GRB 090510}, the Fermi collaboration has not published the
high energy afterglow data yet. According to \citet{Giuliani09}, the
high energy photon flux recorded by \emph{AGILE} is $\sim
0.01(t/2~{\rm s})^{-1.3}~{\rm ph~cm^{-2}~s^{-1}}$. In the IC
scattering model, the number of the seed photons is needed to be
$N_{\rm se}\sim 10^{65}$. Even all seed photons are in near infrared
band ($\sim$ 1 eV), the total energy should be $10^{53}$ erg,
seeming unreasonably large for GRB 090510.

\subsection{Synchrotron radiation of forward shock electrons?}
The spectrum of the synchrotron radiation of shocked electrons can
extend to an energy $\sim 30 {\cal A}\Gamma/(1+z)~{\rm MeV}$
\citep[e.g.,][]{Chengwei96}, where $\Gamma$ is the bulk Lorentz
factor of the emitting region and ${\cal A} \sim (1,~2\pi)$,
depending on the comoving acceleration timescale of the particles.
But usually the IC scattering plays a more important role in
producing high energy afterglow emission. The situation changed in
GRB 080319B, the naked-eye burst with abundant optical and X-ray
afterglow data. With the well constrained parameters, Zou et al.
(2009, Figure 3 therein) have shown that the forward shock
synchrotron radiation dominates over the synchrotron self-Compton
radiation up to an energy $\sim 10$ GeV. The detection prospect for
LAT is pretty good. Our estimated forward shock parameters of GRB
080916C are similar to those of GRB 080319B, a strong forward shock
synchrotron GeV emission is naturally expected \citep[see
also][]{Kumar09}.

In the synchrotron radiation model, the random Lorentz factor of
electrons emitting $\geq 100$ MeV afterglow photons is so high that
$\eta_{_{\rm KN}}\ll 1$ \citep[e.g.,][]{Fan06a}, one should take
$Y\sim 0$ in calculating $\nu_{\rm c}$ otherwise the radiation flux
will be underestimated. For {GRB 080916C}, at $t\sim 400$ s
$\nu_{\rm c}<100 ~{\rm MeV}$, the flux $F_{100~{\rm MeV}}
=F_{\nu,\rm max} (\nu_{\rm c}/\nu_{\rm m})^{-(p-1)/2}({100 \rm
MeV}/h\nu_{\rm c})^{-p/2}\sim 2.7\times 10^{-8}~ {\rm Jy}~ E_{\rm k,
55.6}^{1.05}\epsilon_{\rm B, -2.6}^{0.05}\epsilon_{\rm
e,-1}^{1.2}t_{2.6}^{-1.15}D_{\rm L, 29.1}^{-2}$ and the
corresponding energy flux is $\sim 6.5\times 10^{-9}~{\rm erg
~cm^{-2}~s^{-1}}$, matching the observation $\sim 1.2\times
10^{-9}~{\rm erg ~cm^{-2}~s^{-1}}$. For {GRB 090510}, at $t\sim 5$
s, $h\nu_{\rm c}\sim 18 ~\rm MeV$, so the high energy flux $F_{\rm
100~{MeV}} = F_{\nu,\rm max} (\nu_{\rm c}/\nu_{\rm
m})^{-(p-1)/2}({100 \rm MeV}/h\nu_{\rm c})^{-p/2}\sim
2.0\times10^{-6}~ {\rm Jy}~ \epsilon_{\rm e,-0.1}E_{\rm
k,54}t_{0.7}^{-1}D_{L,28.26}^{-2}$. The corresponding energy flux is
$ \sim 5.0 \times10^{-7} ~{\rm erg~ cm^{-2}~s^{-1}}$. The GeV photon
flux recorded by \emph{AGILE} is $\sim 4\times 10^{-3}~ {\rm ph~
cm^{-2}~s^{-1}}$ for $t\sim 5$ s \citep[see Figure 3
of][]{Giuliani09}, suggesting an energy flux $\sim 6.4\times 10^{-7}
\rm erg~ cm^{-2}~s^{-1}$. So the observation may be accounted for.

As shown in \citet{Zou09}, the synchrotron self-Compton radiation of
such energetic forward shock peaks at TeV energies and may be idea
target for the ground-based Cherenkov telescopes, like MAGIC (Major
Atmospheric Gamma-ray Imaging Cherenkov Telescope) and H.E.S.S. (The
High Energy Stereoscopic System).

\section{Conclusion and Discussion}
In this work we have interpreted the high energy emission and the
afterglow of GRBs 080916C and 090510. For the prompt high energy
emission of GRB 080916C with a featureless Band spectrum, the
standard/unmagnetized internal shock model is disfavored. The main
reason is that in such a model, the fast shells move with a very
high bulk Lorentz factor ($\sim 5\times 10^{3}$) and the thermal
radiation from their photospheres will be too strong to be hidden by
the non-thermal emission of the internal shocks. As for the idea
that the prompt GeV photons are the synchrotron radiation of the
forward shock electrons, we predict very unusual X-ray (for
$t<10^{3}$ s) and optical (for $t<1$ day) afterglow light curves.
The lack of early  afterglow observation, however, hampers us to
test the model. If the prompt GeV photons and the soft gamma-rays
are from the same region, a non-baryonic component seems needed
\citep{ZP09,Fan09}. For GRB 090510, the prompt spectrum consists of
two distinct components. The MeV emission may be from the
photosphere while the GeV emission is produced in the IC scattering
of the photosphere photons by the shocked electrons. We suggest that
the outflow of GRB 090510 is baryonic.

The circum-burst medium of GRBs 080916C and 090510 is wind-like and
ISM, respectively. The standard external shock model can reproduce
the afterglow data reasonably well. The common features are the low
density of the medium they are expanding into and the very high
isotropic-equivalent kinetic energy of the outflows.
We have proposed a simple method to estimate the total number of the
seed photons supposing the GeV afterglow emission is due to the IC
radiation of the forward shock electrons. Such a model is disfavored
because the seed photons needed in the modeling are too many to be
realistic. Though other possibilities, for example the GeV afterglow
photons are the synchrotron self-Compton radiation of the extended
X-ray emission, cannot be ruled out, the high-energy afterglow
detected in these two bursts may be just the synchrotron radiation
of the forward shock electrons. Our analysis is then in support of
the "prediction" of Zou et al. (2009) in GRB 080319B and the
suggestion of \citet{Kumar09} for GRB 080916C. GRBs 080319B, 080916C
and 090510 are very unusual. They are extremely bright\footnote{For
the two long bursts $E_{\rm \gamma}>10^{54}$ erg, while for the
short burst GRB 090510 $E_{\rm \gamma}>10^{53}$ erg. All are at
least 1 order of magnitude brighter than the normal long and short
GRBs.} and may have a very large initial bulk Lorentz factor. Both
facts are helpful to give rise to a strong GeV synchrotron radiation
of the forward shock. The number density of the circum-burst medium
is very low, which lowers the detection prospect of the IC radiation
component for LAT. For normal GRBs, the detection prospect of the
GeV synchrotron radiation of the forward shock will be much less
promising.

\section*{Acknowledgments}
We are grateful to R. Margutti for providing XRT data and P. Kuin
for communication. This work was supported in part by the National
Natural Science Foundation of China under grant 10603003 (for
W.H.G.),  the Danish National Science Foundation, Chinese Academy of
Sciences, and National basic research program of China under grant
2009CB824800 (for Y.Z.F.).

\clearpage


\begin{thebibliography}{}
\bibitem[Abdo et al. (2009a)]{Abdo09} Abdo, A. et al. 2009a, Science, 323,1688
\bibitem[Abdo et al. (2009b)]{Abdo09b} Abdo, A. et al. 2009b, Nature submitted (arXiv:0908.1832)
\bibitem[Cheng \& Wei(1996)]{Chengwei96} Cheng, K.S., \& Wei, D.M., 1996, MNRAS, 283, L133
\bibitem[Chevalier \& Li(2000)]{Chevalier00} Chevalier,R.A., \& Li, Z.Y., 2000, ApJ, 536, 195
\bibitem[De Pasquale et al.(2009)]{Pasquale09} De Pasquale, M., et al., 2009, ApJ submitted(arXiv:0910.1629)
\bibitem[Fan (2009)]{Fan09} Fan, Y. Z., 2009, MNRAS submitted
\bibitem[Fan \& Piran(2008)]{Fan08} Fan, Y.Z., \& Piran, T., 2008, Front. Phys. Chin., 3, 306
\bibitem[Fan \& Piran(2006a)]{Fan06a} Fan, Y.Z., \& Piran, T., 2006a, MNRAS, 369, 197
\bibitem[Fan \& Piran(2006b)]{Fan06} Fan, Y.Z., \& Piran, T., 2006b, MNRAS, 370, L24
\bibitem[Gao(2009)]{Gao09} Gao, W.H., 2009, ApJ, 697, 1044
\bibitem[Giuliani et al.(2009)]{Giuliani09} Giuliani, A., et al. 2009, ApJL in press (arXiv:0908.1908)
\bibitem[Greiner et al.(2009a)]{Greiner09a} Greiner, J., et al. 2009a, ApJ, 693, 1912
\bibitem[Greiner et al.(2009b)]{Greiner09} Greiner, J., et al. 2009b,
A\&A, 498, 89
\bibitem[Kumar \& Barniol Duran(2009)]{Kumar09} Kumar, P., \& Barniol Duran, R., 2009, MNRAS, in press (arXiv:0905.2417)
\bibitem[Paczy\'nski (1990)]{Pacz90} Paczynski B., 1990, ApJ, 363, 218
\bibitem[Piran(1999)]{Piran99} Piran, T., 1999, Phys. Rep., 314, 575
\bibitem[Sari et al.(1996)]{Sari96} Sari, R., Narayan, R., \& Piran, T., 1996, ApJ, 473,
204
\bibitem[Xue et al. (2009)]{Xue09} Xue, R. R., Fan, Y. Z., \& Wei, D. M., 2009, A\&A, 498, 671
\bibitem[Yost et al.(2003)]{Yost03} Yost, S.A., et al. 2003, ApJ, 597, 459
\bibitem[Zhang \& M\'esz\'aros(2004)]{ZM04} Zhang, B., \& M\'esz\'aros,
P., 2004, Int. J. Mod. Phys. A, 19, 2385
\bibitem[Zhang \& Pe'er(2009)]{ZP09} Zhang, B., \& Pe'er, A., 2009, ApJ, 700, L65
\bibitem[Zou et al.(2009)]{Zou09} Zou, Y.C., Fan, Y.Z., \& Piran, T., 2009, MNRAS, 396, 1163
\end{thebibliography}
\end{document}